\documentclass[aps,twocolumn,10pt]{revtex4}
\usepackage{amsmath,graphicx,amssymb}
\usepackage{float}
\usepackage[caption = false]{subfig}

\begin{document}
\title{Orbital order, spin waves, and doping effects in the $(\pi,0)$ SDW state \\
 -- comparative study of the three band models for iron pnictides}
\author{Nimisha Raghuvanshi}
\email{nimishar@iitk.ac.in}
\author{Avinash Singh}
\email{avinas@iitk.ac.in}
\affiliation{Department of Physics, Indian Institute of Technology Kanpur, Kanpur - 208016, India}

\begin{abstract}
Aimed at identifying the role of microscopic Hamiltonian parameters on 
spin wave excitations, orbital order, and magnetic moments in the $(\pi,0)$ SDW state, two different three band models for iron pnictides are compared --- one at two-third filling ($n=4$) and another at half filling $(n=3)$ --- both of which yield qualitatively correct Fermi surface structure in the paramagnetic state. Spin wave analysis of the model at $n=4$ shows instability of the SDW state, which is attributed to weakly developed magnetic moments and weak magnetic couplings due to overfilling, as inferred from the observed stabilization of the SDW state and enhancement of magnetic excitation energies upon hole doping. In contrast, the model at $n=3$ (half filling) yields a gapped SDW state with well developed magnetic moments and the calculated spin wave excitations are in excellent agreement with INS experiments. The sign of the orbital order ($n_{xz} - n_{yz} \approx +0.2$) is also in agreement with experiments. Both the zone boundary spin wave energies in the F and AF directions as well as the orbital order are shown to peak near half filling, highlighting the correlation between orbital order and SDW state stabilization. 

\end{abstract}

\maketitle

\section{Introduction}

The complex multi-orbital microscopic description of iron pnictides is qualitatively well established by density functional theory (DFT) calculations \cite{lebegue_2007,singh_2008,cao_2008,xu_2008,zhang_2009} 
and angle-resolved photoemission spectroscopy (ARPES) experiments \cite{liu_2009,lu_2009,zhang_2011}. 
The Fermi surface, 
with two nearly circular hole pockets around $\Gamma$ $(0,0)$ and elliptical electron pockets each around X $(\pm \pi,0)$ and Y $(0,\pm \pi)$ points in the paramagnetic (PM) state, 
is mainly composed of $d_{xz}$, $d_{yz}$ and $d_{xy}$ Fe orbitals \cite{yi_2009,kondo_2010,yi_2011,brouet_2012,kordyuk_2013}.
The two hole pockets involve mixing between the $xz$ and $yz$ orbitals, while the electron pockets are composed of $xy$ and $xz/yz$ orbitals.
In addition to the complex Fermi surface reconstruction \cite{yi_2009,yi_2011} 
through the transition from paramagnetic to antiferromagnetic $(\pi,0,\pi)$ state below $T_N \approx 200$ K \cite{goldman_2008}, 
sharp spin-wave excitations on an energy scale $\sim 200$ meV are reported by inelastic neutron scattering (INS) measurements
\cite{zhao_2009,diallo_2009,ewings_2011,harriger_2011}.

Another extensively discussed experimental finding in iron pnictides is the ferro orbital order between the $d_{xz}$ and $d_{yz}$ Fe orbitals 
\cite{yi_2011,fuglsang_2011,kim_2013,yi_2014}.
ARPES and x-ray linear dichroism (XLD) measurements have revealed a relative shift between the $xz$ and $yz$ bands in iron pnictides, 
with the $xz$ orbital lying slightly lower and therefore slightly more filled. 
The observation of spin wave excitations persisting even above the N\'{e}el temperature \cite{diallo_2010,ewings_2011,harriger_2011,harriger_2012,lu_2014}, 
indicates that short range antiferromagnetic (AF) and ferromagnetic (F) order remain in the $a$ and $b$ directions, respectively, even above the disordering temperature. 
This may account for the narrow nematic phase \cite{kasahara_2012,zhou_2013,lee_2009,kruger_2009,lv_2009,chen_2010} above the N\'{e}el temperature where ferro orbital order \cite{yi_2011,kim_2013} and structural distortion survive, 
as well as the temperature dependence of the measured anisotropies in $a$ and $b$ directions of magnetic excitations and resistivity \cite{chu_2010,yi_2011,chu_2012,lu_2014,fernandes_2014}.

In view of above results, three orbital tight binding models involving the $d_{xz}$, $d_{yz}$ and $d_{xy}$ Fe $3d$ orbitals have been widely studied in order to gain insight into the magnetic and superconducting state of iron pnictides 
\cite{lee_2008,kruger_2009,yu_2009,zhou_2010,maria_2010,maria_1_2010,daghofer_2012,ghosh_2014_a}.
In particular, the proposed three orbital model at two-third filling $(n=4)$ \cite{maria_2010} produces the Fermi surface in agreement with the DFT calculations and ARPES experiments.
Similarly, the three band model at half filling $(n=3)$\cite{ghosh_2014_a} also produces the correct Fermi surface, with the orbital composition of the electron and hole pockets agreeing well with more realistic five-band models \cite{kuroki_2008,graser_2009,graser_2010} and ARPES results. 

The interplay between the complex multi-orbital character of the microscopic description of iron pnictides and their macroscopic electronic, magnetic, and orbital properties is therefore of strong current interest. In this work, we will investigate the $(\pi,0)$ spin density wave (SDW) state of these two models in detail, and analyze the spin wave spectral function, orbital resolved magnetization, and orbital order. 

This article is organized as follows.
Section ${\rm II}$ starts by introducing the tight binding model and the interaction terms for the three band model. 
The Hartree-Fock (HF) level Hamiltonian in the spontaneously-broken-symmetry $(\pi,0)$ ordered magnetic state is presented, and the numerical scheme for evaluating the transverse spin fluctuation propagator and the spin wave spectral function in the random phase approximation (RPA) are briefly discussed.
Sections ${\rm III}$ and ${\rm IV}$ describe the results of the spin wave spectral function, orbital resolved magnetization, and orbital order for the two three band models of Refs. \cite{maria_2010} (at two-third filling) and \cite{ghosh_2014_a} (at half filling). Conclusions based on this comparative study are presented in Section ${\rm V}$, aimed at identifying the favorable multi-band model features with regard to spin wave excitations and orbital order.

\section{Three orbital model for iron pnictides}

The tight binding part of the three orbital model involving $d_{xz}$, $d_{yz}$ and $d_{xy}$ Fe $3d$ orbitals is

\begin{equation} \label{eq:1}
 H_{\rm t} = -\sum_{\langle i j \rangle \mu\nu\sigma}  t_{i j}^{\mu\nu} (a_{i \mu \sigma}^{\dagger} a_{j \nu \sigma} + a_{j \nu \sigma}^{\dagger} a_{i \mu \sigma}) + \varepsilon_{xy} \sum_{i} n_{i \mu=xy}
\end{equation}
where $\langle i j \rangle$ refers to the nearest-neighbor (NN) and next-nearest-neighbor (NNN) pairs of lattice sites, $\mu, \nu$ represent the $d_{xz}, d_{yz}, d_{xy}$ orbitals, $t_{ij}^{\mu\nu}$ represent the corresponding hopping terms, and $\varepsilon_{xy}$ is the energy offset of the $xy$ orbital relative to the degenerate $xz/yz$ orbitals. We will consider the following band energies corresponding to hopping terms discussed below:

\begin{equation} \label{eq:2}
\begin{split}
\varepsilon_{\bf k}^{xz} &= -2{t}_1\cos{k}_x-2{t}_2\cos{k}_y-4{t}_3\cos{k}_x\cos{k}_y\\
                         &= \varepsilon_{\bf k}^{1x}+\varepsilon_{\bf k}^{2y}+\varepsilon_{\bf k}^{3}\\
\varepsilon_{\bf k}^{yz} &= -2{t}_1\cos{k}_y-2{t}_2\cos{k}_x-4{t}_3\cos{k}_x\cos{k}_y\\
                     &= \varepsilon_{\bf k}^{1y}+\varepsilon_{\bf k}^{2x}+\varepsilon_{\bf k}^{3}\\
\varepsilon_{\bf k}^{xy} &= -2{t}_5(\cos{k}_x+\cos{k}_y)-4{t}_6\cos{k}_x\cos{k}_y+\varepsilon_{\rm off}\\
                     &= \varepsilon_{\bf k}^{5x}+\varepsilon_{\bf k}^{5y}+\varepsilon_{\bf k}^{6}+\varepsilon_{xy}\\
\varepsilon_{\bf k}^{xz,yz} &= \varepsilon_{k}^{yz,xz} =-4{t}_4\sin{k}_x\sin{k}_y= \varepsilon_{\bf k}^{4}\\
\varepsilon_{\bf k}^{xz,xy} &= \overline{\varepsilon_{k}^{xy,xz}}= -2i{t}_7\sin{k}_x-4i{t}_8\sin{k}_x\cos{k}_y\\
                            &= \varepsilon_{\bf k}^{7x}+ \varepsilon_{\bf k}^{8x}\\
\varepsilon_{\bf k}^{yz,xy} &= \overline{\varepsilon_{k}^{xy,yz}} = -2i{t}_7\sin{k}_y-4i{t}_8\cos{k}_x\sin{k}_y\\
                            &= \varepsilon_{\bf k}^{7y}+ \varepsilon_{\bf k}^{8y}\\
\end{split}
\end{equation}

Within the degenerate $d_{xz}$, $d_{yz}$ sector, the hopping parameters $t_1$ and $t_2$ correspond to NN $\sigma$ and $\pi$ bonds linking similar orbitals, while NNN hoppings $t_3$ and $t_4$ represent overlap between similar and different orbitals, respectively. The hopping parameters $t_5$ and $t_6$ are the NN and NNN hoppings for the third orbital, $d_{xy}$. The $d_{xz} / d_{yz}$ orbitals hybridize with the $d_{xy}$ orbital via the $t_7$ (NN) and $t_8$ (NNN) mixing terms. The values of the hopping parameters for the two different models  considered are given in sections III and IV.  

The interaction part of the Hamiltonian is given by:

\begin{equation} \label{eq:3}
\begin{split}
 H_{\rm int} =     U \sum_{ i \mu}  n_{i \mu \uparrow} n_{i \mu \downarrow}
             - 2 J \sum_{ i, \mu \neq \nu } {{\bf{S}}_{i \mu }} \cdot {{\bf{S}}_{i \nu }}\\
             + (U'-J/2) \sum_{i, \mu \neq \nu } n_{ i \mu} n_{ i \nu}\\
             + J' \sum_{i, \mu \neq \nu } ( a_{i \mu \uparrow }^{\dagger} a_{i \mu \downarrow }^{\dagger} a_{i \nu \downarrow } a_{i \nu \uparrow } + \text{H.c.})
\end{split}
\end{equation}
Here $U$ is the intra-orbital Coulomb interaction, $J$ is the Hund's rule coupling term, and $U'-J/2$ (where $U'=U-2J$) represents the inter-orbital density interaction term \cite{maria_2010}. 
The inter-orbital pair hopping term $J'$ is equal to Hund's rule coupling term $J$ by symmetry. For the subsequent discussion, we have fixed the ratio $J/U \approx 1/4$. 

We now consider the $(\pi,0)$ ordered SDW state of the three band model. At the Hartree-Fock level, the Hamiltonian matrix in the composite three-orbital ($xz$ $yz$ $xy$), two-sublattice (A B) basis (A$xz$ A$yz$ A$xy$ B$xz$ B$yz$ B$xy$) is obtained as: 

\onecolumngrid
\begin{equation}
H_{\rm HF}^{\sigma} ({\bf k}) = \left [ \begin{array}{cccccc} -\sigma
\Delta_{xz} + \varepsilon_{\bf k}^{2y}   & 0 & 0 & \varepsilon_{\bf k}^{1x}
+ \varepsilon_{\bf k}^{3} & \varepsilon_{\bf k}^{4} & \varepsilon_{\bf
k}^{7x}
+ \varepsilon_{\bf k}^{8x}
\\
0 & -\sigma \Delta_{yz} + \varepsilon_{\bf k}^{1y} & \varepsilon_{\bf
k}^{7y}  &
\varepsilon_{\bf k}^{4} & \varepsilon_{\bf k}^{2x} + \varepsilon_{\bf
k}^{3}  &
\varepsilon_{\bf k}^{8y} \\
0 & -\varepsilon_{\bf k}^{7y} & - \sigma \Delta_{xy} + \varepsilon_{\bf
k}^{5y} + {\varepsilon}_{xy} & - \varepsilon_{\bf k}^{7x}-
\varepsilon_{\bf
k}^{8x} & -\varepsilon_{\bf k}^{8y} & \varepsilon_{\bf k}^{5x} +
\varepsilon_{\bf k}^{6} \\
\varepsilon_{\bf k}^{1x} + \varepsilon_{\bf k}^{3} & \varepsilon_{\bf
k}^{4} &
\varepsilon_{\bf k}^{7x} + \varepsilon_{\bf k}^{8x} & \sigma \Delta_{xz} +
\varepsilon_{\bf k}^{2y} & 0 & 0 \\
\varepsilon_{\bf k}^{4} & \varepsilon_{\bf k}^{2x} + \varepsilon_{\bf
k}^{3} &
\varepsilon_{\bf k}^{8y} & 0 & \sigma \Delta_{yz} + \varepsilon_{\bf
k}^{1y} &
\varepsilon_{\bf k}^{7y} \\
- \varepsilon_{\bf k}^{7x} - \varepsilon_{\bf k}^{8x} & -\varepsilon_{\bf
k}^{8y} & \varepsilon_{\bf k}^{5x} + \varepsilon_{\bf k}^{6} & 0 &
- \varepsilon_{\bf k}^{7y} & \sigma
\Delta_{xy} + \varepsilon_{\bf k}^{5y} + {\varepsilon}_{xy}  \\
\end{array}
\right ]
\label{eq:7}
\end{equation}

\twocolumngrid

Here the exchange field for orbital $\mu$ is given by:
\begin{equation}\label{eq:4}
\begin{split}
 2\Delta_\mu = U m_{\mu} + J\sum_{\nu \neq \mu} m_{\nu}
\end{split}
\end{equation}
in terms of the sublattice magnetization $m_{\mu} = n_{\mu}^{\uparrow} - n_{\mu}^{\downarrow}$, where $n_{\mu}^{\sigma}$ represents the electronic density for spin $\sigma$ and orbital $\mu$. The density part of the interaction Hamiltonian, $(5J-U)(n_{\mu} - n_{\nu})/2$, yields small contribution for the parameters considered ($J/U \approx 1/4$), and is therefore neglected in the following discussion.

In the above $(\pi,0)$ SDW state $(|\Psi_0 \rangle)$, we consider the time-ordered transverse spin fluctuation propagator in the orbital-sublattice basis:
\begin{equation}
\begin{split}
\chi^{-+}({\bf q},\omega) = \int dt \sum_i e^{i\omega(t-t')} 
e^{-i{\bf q}.({\bf r}_i -{\bf r}_j)} \\ 
\times  \langle \Psi_0 | T [ S_{i\mu} ^- (t) S_{j\nu} ^+ (t')]|\Psi_0 \rangle \;
\end{split}
\end{equation}
involving the spin lowering and raising operators. In the RPA, the spin fluctuation propagator is obtained as:
\begin{equation}\label{eq:5}
[\chi^{-+} _{\rm RPA} ({\bf q},\omega)] = \frac{[\chi^0 ({\bf q},\omega)]}{{\bf 1} - [U][\chi^0 ({\bf q},\omega)]}
\end{equation}
where the local interaction matrix $[U]$ in the orbital-sublattice basis includes the diagonal $U$ terms and the off-diagonal $J$ terms with respect to orbitals. The bare particle-hole propagator:
\begin{equation}
\begin{split}
[\chi^0 ({\bf q},& \omega)]_{a b} =  \sum_{k, l, m} \left[{\frac{{\phi^{a*}_{{\bf k} \uparrow l}}{\phi^{b} _{{\bf k} \uparrow l}}{\phi^{a} _{{\bf {k-q}} \downarrow m}}{\phi^{b*} _{{\bf {k-q}} \downarrow m}}}{E^+_{{\bf {k-q}} \downarrow m} - E^-_{{\bf k} \uparrow l}  + \omega - i \eta}} \right. \\
 & \left. + {\frac{{\phi^{a*}_{{\bf k} \uparrow l}}{\phi^{b} _{{\bf k} \uparrow l}}{\phi^{a} _{{\bf {k-q}} \downarrow m}}{\phi^{b*} _{{\bf {k-q}} \downarrow m}}}{E^+_{{\bf k} \uparrow l} - E^-_{{\bf {k-q}} \downarrow m}  - \omega - i \eta}}\right]
\end{split}
\end{equation}
is evaluated in the orbital-sublattice basis by integrating out the fermions in the $(\pi, 0)$ ordered SDW state.
Here $E_{{\bf k} \sigma}$ and $\phi_{{\bf k} \sigma}$ are the eigenvalues and eigenvectors of the Hermitian Hamiltonian matrix (\ref{eq:7}), the orbital-sublattice basis indices $a$, $b$ run through 1-6, and $l$, $m$ indicate the six eigenvalue branches.
The superscripts + (-) refer to particle (hole) energies above (below) the Fermi energy. It should be noted that eigenvectors $\phi_{{\bf k} \sigma}$ are complex due to imaginary hopping terms $t_7$ and $t_8$ in Eq.~\ref{eq:2}. Without the $\eta$ term, the $[\chi^0 ({\bf q},\omega)]$ matrix is Hermitian.
The results in SDW state are obtained for a $100 \times 100$ mesh in $\bf k$ space and a finite damping $\eta$ of 5meV. 

The spin wave spectral function was obtained by taking the trace of the imaginary part of the spin fluctuation propagator matrix: 
\begin{equation}\label{eq:6}
A _{\bf q} (\omega) = \frac{1}{\pi} {\text {Tr} \; \text {Im} [\chi^{-+} _{\rm RPA} ({\bf q},\omega)]}
\end{equation}

\begin{figure}[h]
    \centering
     \includegraphics[width=2.0in]{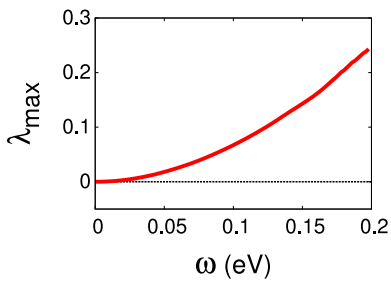}
       \caption{The largest eigenvalue $\lambda_{\rm max}(q=0)$ identically vanishes for $\omega=0$, explicitly confirming the existence of the Goldstone mode in the broken symmetry SDW state for the three band model with complex hopping terms.}
    \label{mueig1}%
\end{figure}

In view of the continuous spin rotation symmetry of the Hamiltonian considered, the broken symmetry SDW state must possess the zero energy Goldstone mode. In order to confirm this for the three band model with complex hopping terms, we have evaluated the largest eigenvalue $\lambda_{\rm max}(\omega)$ of the matrix ${[U] - [U][\chi^0 ({\bf q},\omega)][U]}$ (symmetrized denominator in Eq.~\ref{eq:5}) at ${\bf q}=0$. As shown in  Fig.~\ref{mueig1}, $\lambda_{\rm max}(\omega)$ is identically zero for $\omega = 0$, confirming the zero energy pole in the spin wave propagator corresponding to the Goldstone mode.

In the next two sections, we will consider two distinct three-band models. We will focus on the following aspects in the comparative study:
(1) Spin wave excitations in the SDW state, particularly spin wave energy at the ferro zone boundary ($\omega_{\rm FZB}$) corresponding to wave vector ${\bf q}=(0,\pi)$ in the $(\pi,0)$ SDW state. 
(2) Orbital order between the $xz$ and $yz$ orbitals and its relation to hopping anisotropy $(t_1,t_2)$ in the model.
(3) Orbital resolved sublattice magnetizations in the $xz$, $yz$ and $xy$ orbitals.
(4) Doping dependence of sublattice magnetizations and $\omega_{\rm FZB}$.
(5) Band filling of $xy$ orbital and emergent F spin couplings due to exchange of particle-hole propagator as in metallic ferromagnets.
(6) To what extent is the nesting picture alone adequate as far as magnetic excitations are concerned.

\section{Three band model of Ref.~\cite{maria_2010}}

Within the crystal field splitting picture, due to the orientation of Fe $3d$ orbitals ($xz,yz,xy,x^{2}-y^{2},3z^{2}-r^{2}$) and the As atom positions,
the ${x^{2}-y^{2}}$ and ${3z^{2}-r^{2}}$ orbitals are lower in energy relative to the other three ($xz,yz,xy$) orbitals out of which the $xy$ orbital is pushed up highest. 
With $x^{2}-y^{2}$ and $3z^{2}-r^{2}$ orbitals thus doubly occupied, only two electrons (out of the six Fe 3d orbital electrons in undoped iron pnictides) are left for the remaining three orbitals \cite{kruger_2009}.
However, at this one-third filling indicated in the crystal field splitting picture, the Fermi surface of the three band model does not match with local density approximation (LDA) results.
Therefore, the two-third filling case ($n=4$), suggested by band-structure calculations \cite{boeri_2008,haule_2008}, has been mainly considered in Ref.~\cite{maria_2010}, where the three band model hopping parameters considered are given in the table below.

\begin{table}[H]
\label {tab }
\centering
\begin{tabular}{|c|c|c|c|c|c|c|c|c|}
\hline
$t_1$&$t_2$&$t_3$&$t_4$&$t_5$&$t_6$&$t_7$&$t_8$&$\varepsilon_{xy}$\\
\hline
-0.06&-0.02&-0.03&+0.01&-0.2&-0.3&+0.2&-0.1&0.4\\
\hline
\end{tabular}
\end{table}

At two-third filling, the three orbital model yields two circular hole pockets around $\Gamma$ $(0, 0)$ with $xz,yz$ contributions and an elliptical electron pocket each around
X $(\pm \pi, 0)$ and Y $(0, \pm \pi)$ points having contribution from $xy/yz$ and $xy/xz$ orbitals respectively on the Fermi surface.

\begin{figure}[H]
    \centering
    \includegraphics[width=2.5in]{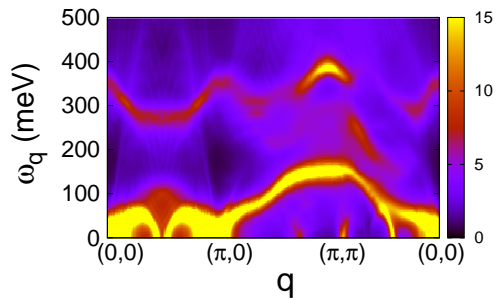}
    \caption{Spin wave spectral function in the $(\pi, 0)$ SDW state of the three band model of Ref.~\cite{maria_2010} 
             showing instability at the AFZB at two-third filling ($n=4$).} 
    \label{m_sus}%
\end{figure}

In this section, we will investigate the spin wave spectral properties in the $(\pi, 0)$ state of the three band model of Ref.~\cite{maria_2010} at the suggested two-third filling for $U=1$ eV. Fig.~\ref{m_sus} shows the calculated spin wave spectral function at two-third filling where the darker color indicates lower intensity. The most significant feature is the negative energy modes observed near the antiferromagnetic zone boundary (AFZB), indicating instability of the SDW state.
The overall spin wave energy scale in the AF direction is also significantly lower in comparison to INS measurements.
Moreover, the spectral weight is shared with an optical branch present at $\sim 300$ meV.
We have confirmed that this high energy branch does not originate from particle-hole excitations by examining the imaginary part of the bare particle-hole propagator $\chi^0$.
Evidently, over-filling results in weakening of the AF spin couplings which are expected to be optimized near half filling.

\begin{figure}[h]
    \centering
    \subfloat{{\includegraphics[width=.48\linewidth]{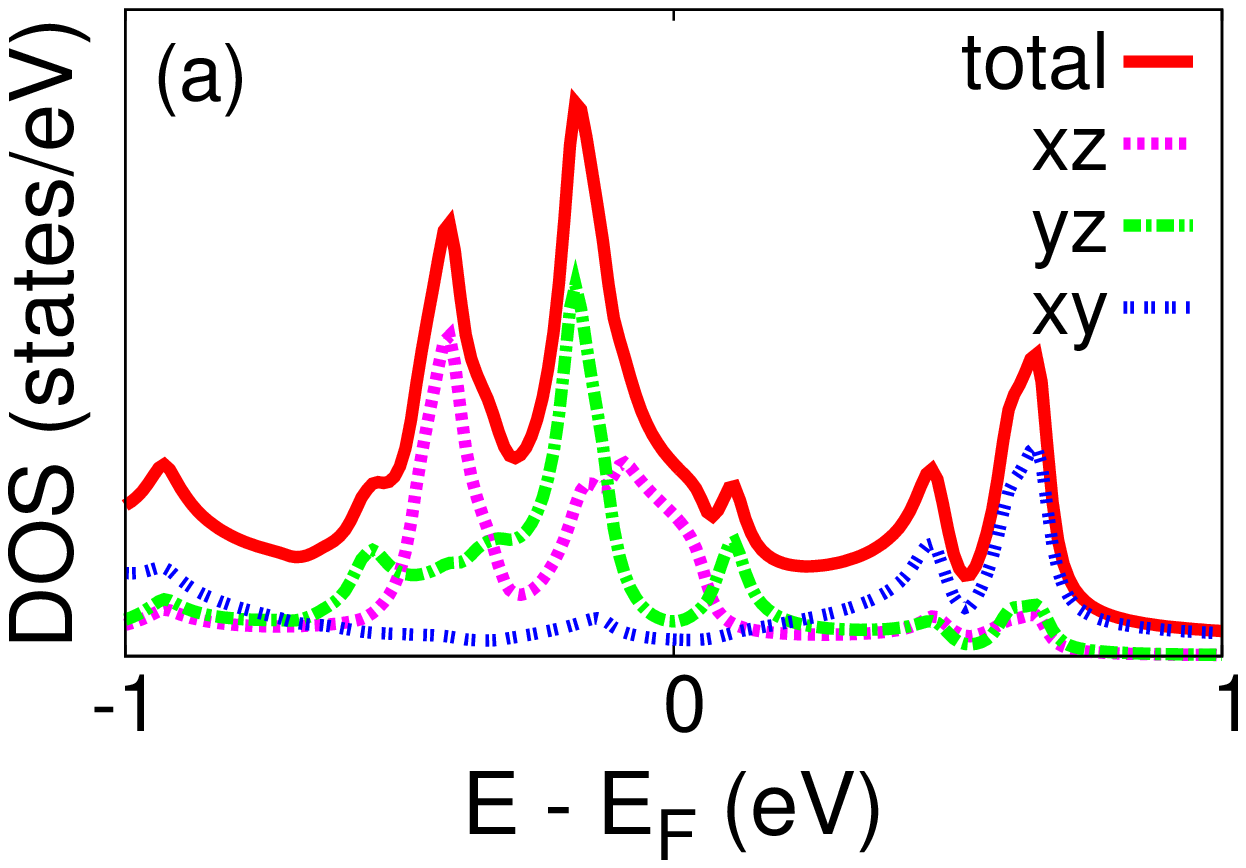}\label{m_doso} }}%
    \qquad
    \hspace{-0.6cm}
    \subfloat{{\includegraphics[width=.47\linewidth]{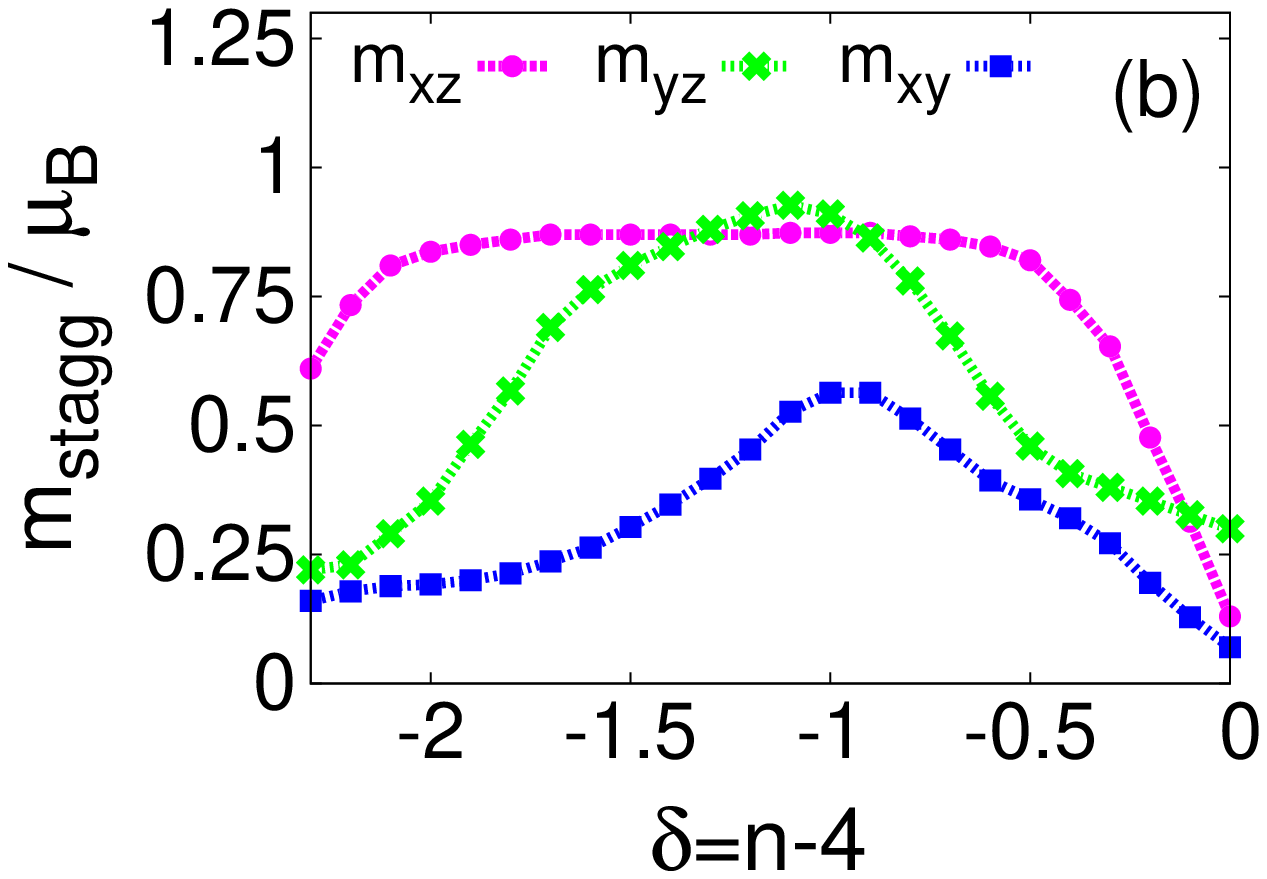}\label{m_mvn} }}%
    \caption{(a) Orbitally resolved density of states at $n=4$ and 
             (b) Orbital resolved staggered magnetization as a function of hole doping with respect to two-third filling
             in the $( \pi, 0)$ SDW state of the three band model of Ref.~\cite{maria_2010}.}
    \label{m_doso_mvn}%
\end{figure}

In the SDW state, DMFT calculations and scanning tunneling microscope (STM) measurements indicate that the Fermi energy lies in the dip of the total density of states (DOS) \cite{yin_2011,chuang_2010}. As seen in Fig.~\ref{m_doso} for the calculated SDW state DOS, this feature is clearly absent in the total DOS for two-third filling. Only a weak dip is seen in the $yz$ orbital DOS at the Fermi energy. 

An important feature of the SDW state near two-third filling in the three band model of Ref.~\cite{maria_2010} is the nearly non-magnetic character of the $xy$ orbital 
($n_{xy}^\uparrow \approx n_{xy}^\downarrow$) due to overfilling.
Hole doping in the SDW state therefore strongly enhances the $xy$ orbital magnetization, as shown in Fig.~\ref{m_mvn}.
The calculated orbital magnetizations $m_{\mu}$ ($\mu= xz,yz,xy$) are in agreement with result of Ref.~\cite{maria_2010} at $n=4$.
However, the magnetic moment in this three band model increases significantly with doping in sharp contrast with the observed reduction of iron moment with doping in neutron scattering experiments \cite{zhao_2008_a}.

\begin{figure}[H]
\centering
\parbox{3cm}{
\hspace{-1.5cm}
\includegraphics[width=0.92\linewidth]{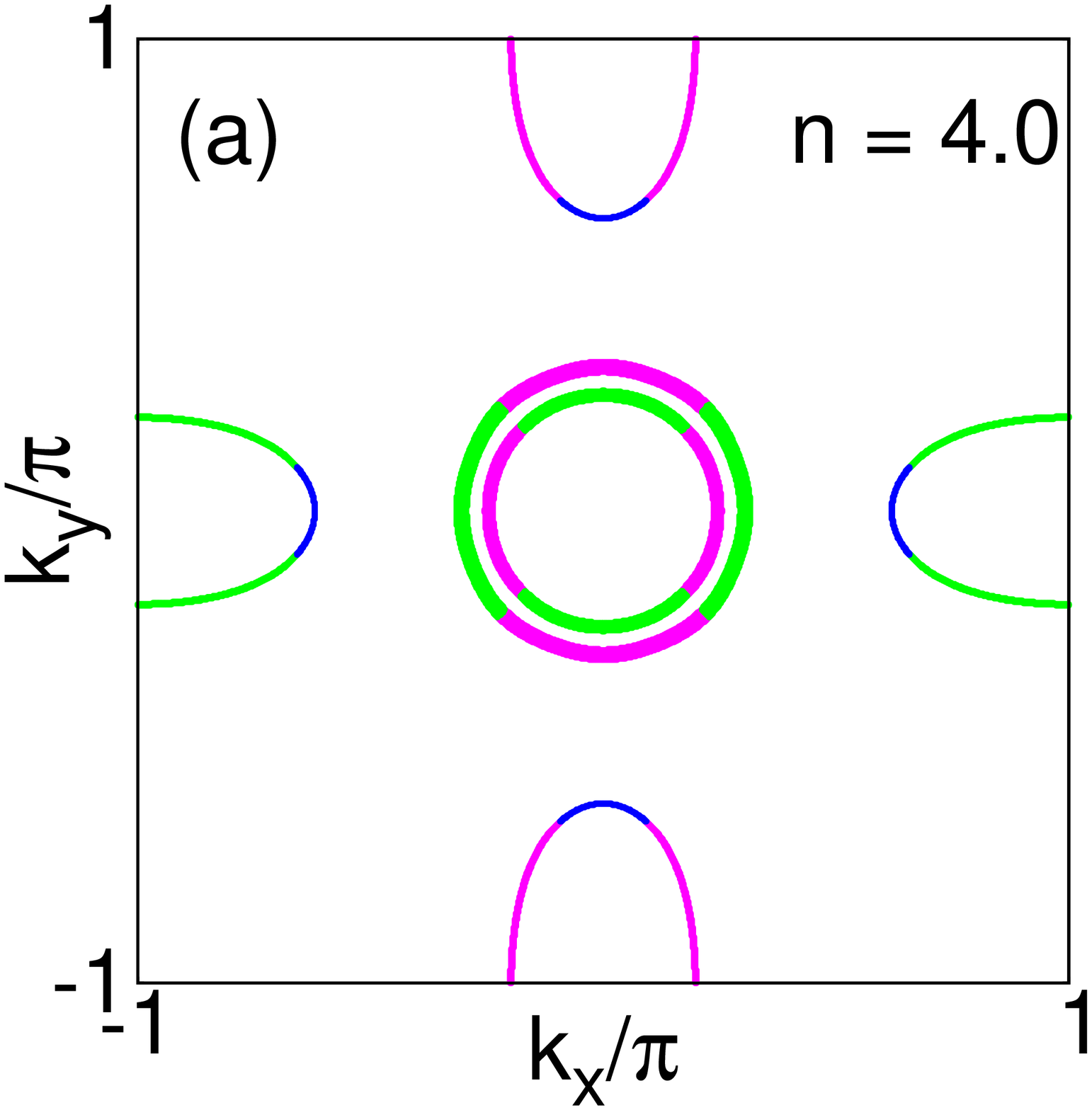}\label{fs1}
\label{fig:2figsB}}
\hspace{-1.0cm}
\begin{minipage}{3cm}
\vspace{-0.1cm}
\includegraphics[width=1.8\linewidth]{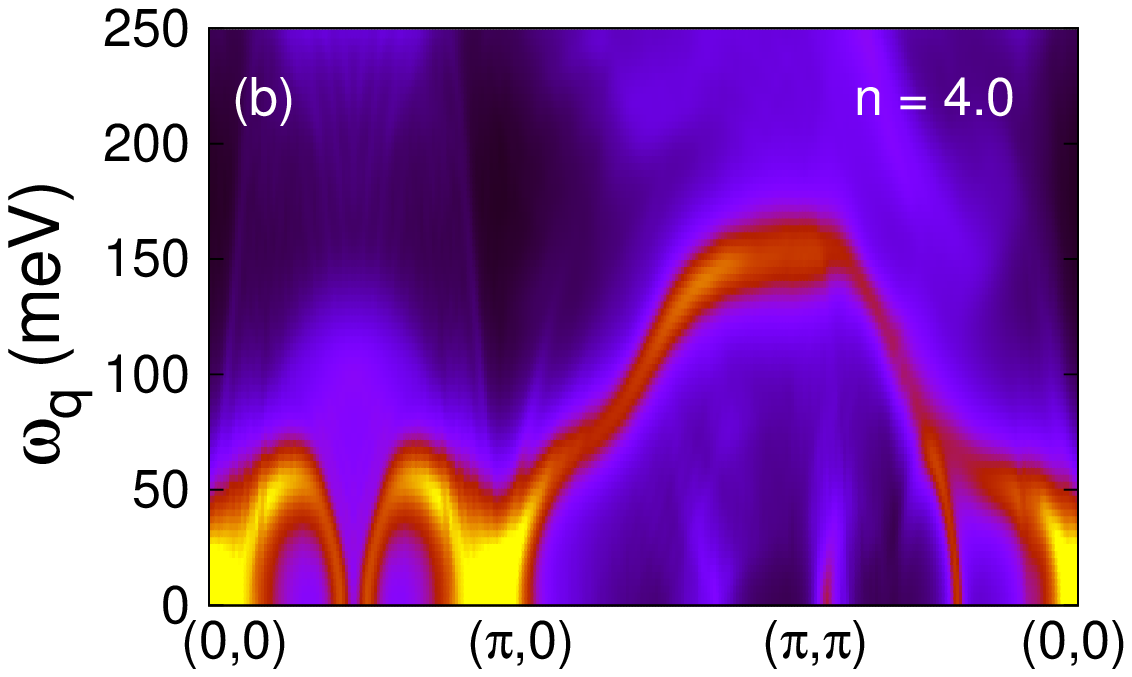}\label{sw1}
\end{minipage}

\vspace{-0.5cm}

\parbox{3cm}{
\hspace{-1.5cm}
\includegraphics[width=0.92\linewidth]{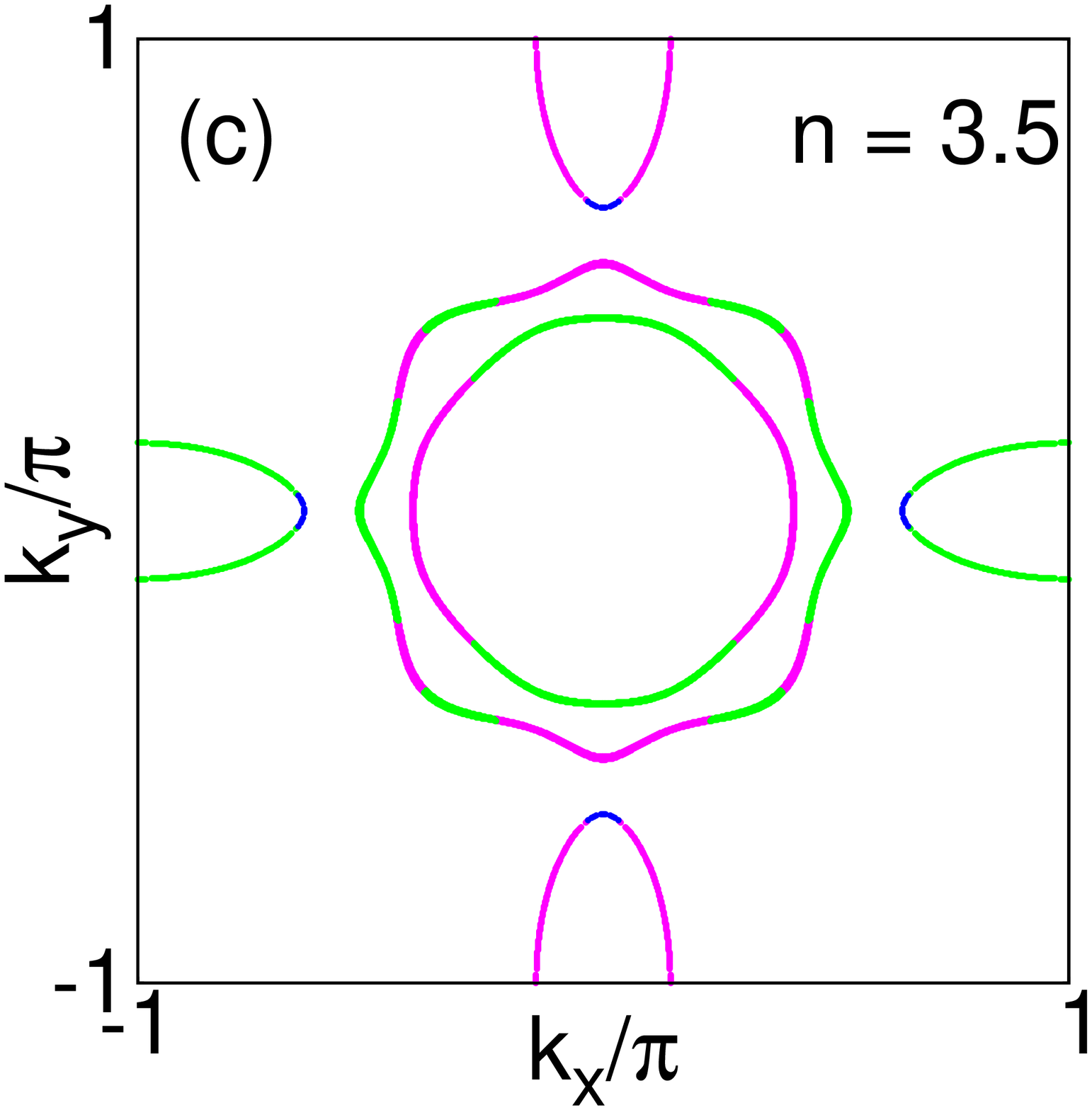}\label{fs2}
\label{fig:2figsC}}
\hspace{-1.0cm}
\begin{minipage}{3cm}
\vspace{-0.1cm}
\includegraphics[width=1.8\linewidth]{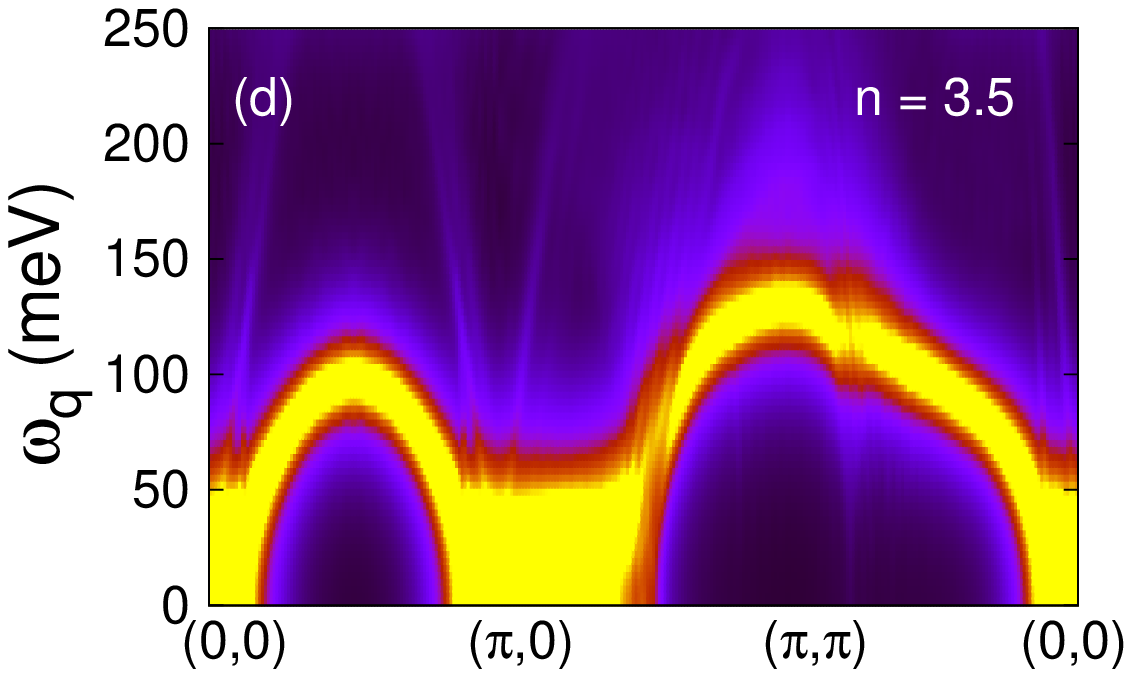}\label{sw2}
\end{minipage}

\vspace{-0.5cm}

\parbox{3cm}{
\hspace{-1.5cm}
\includegraphics[width=0.92\linewidth]{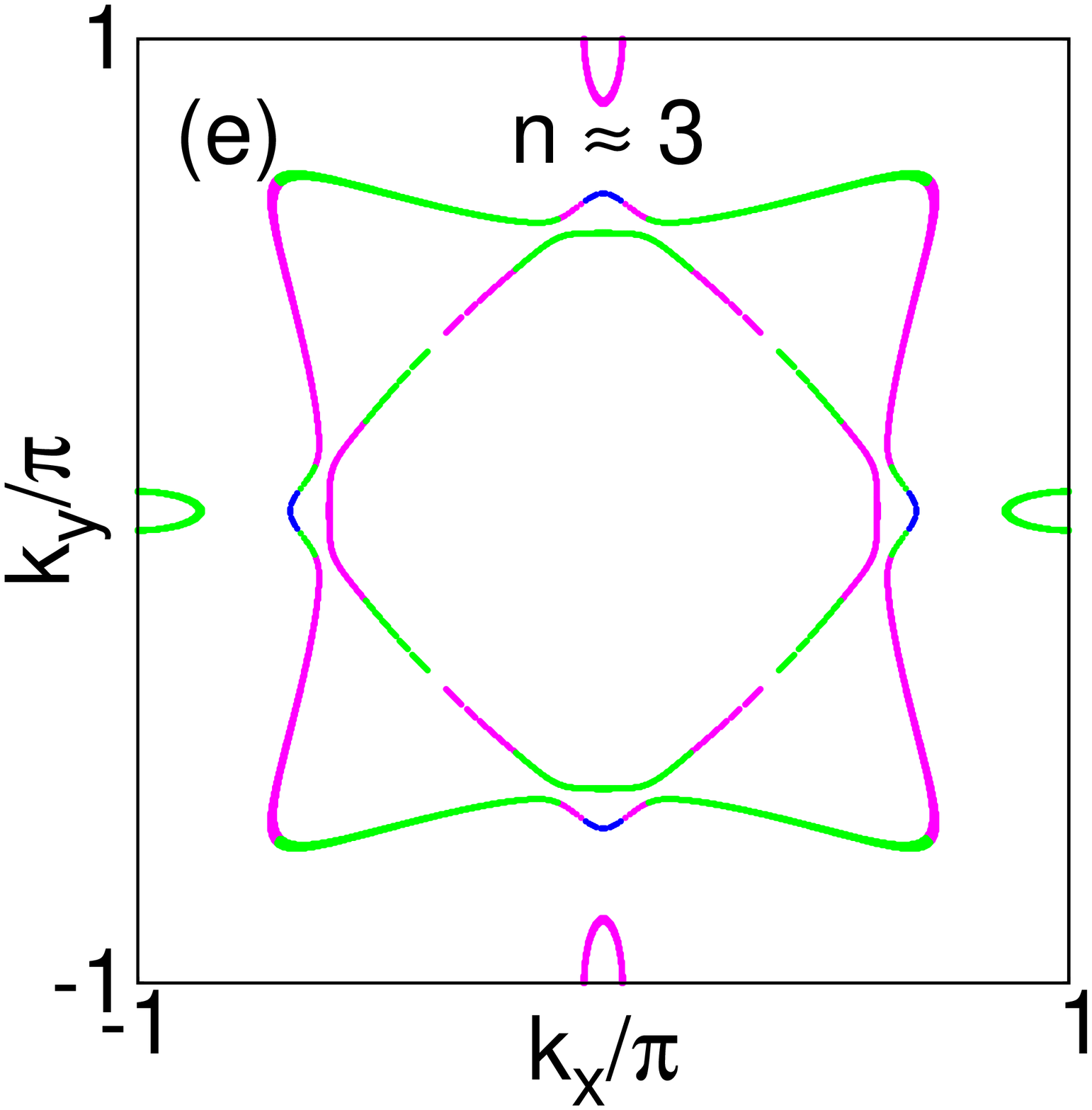}\label{fs3}
\label{fig:2figsD}}
\hspace{-1.0cm}
\begin{minipage}{3cm}
\vspace{-0.1cm}
\includegraphics[width=1.8\linewidth]{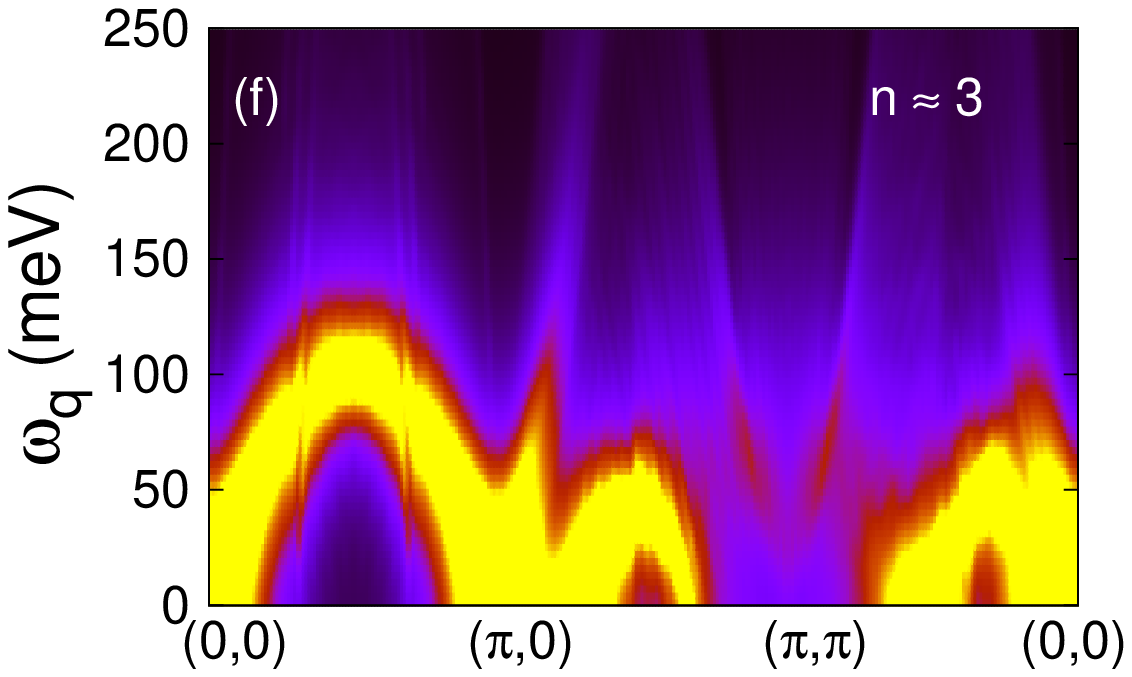}\label{sw3}
\end{minipage}

\vspace{-0.5cm}

\parbox{3cm}{
\hspace{-1.5cm}
\includegraphics[width=0.92\linewidth]{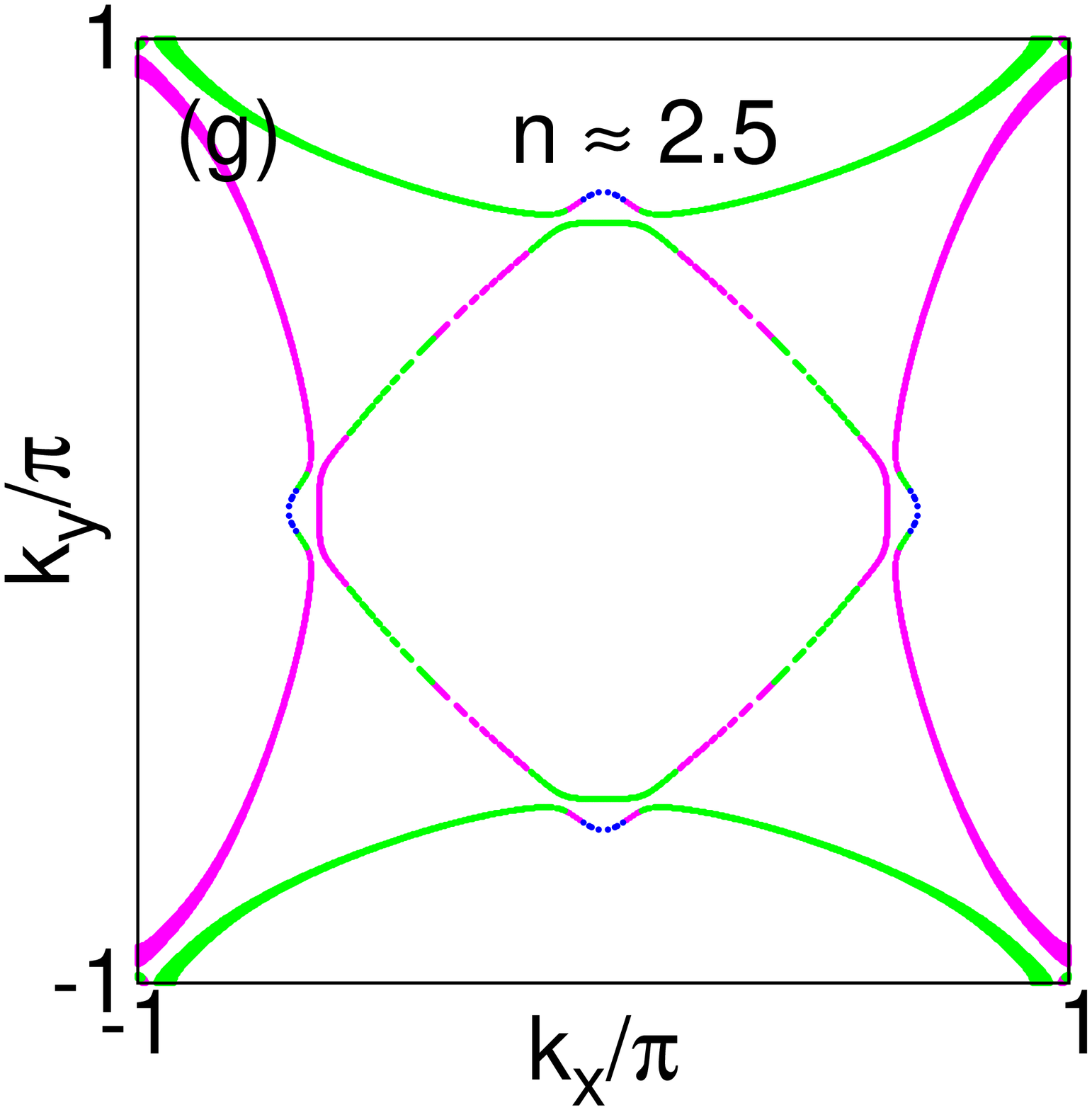}\label{fs4}
\label{fig:2figsE}}
\hspace{-1.0cm}
\begin{minipage}{3cm}
\vspace{-0.1cm}
\includegraphics[width=1.8\linewidth]{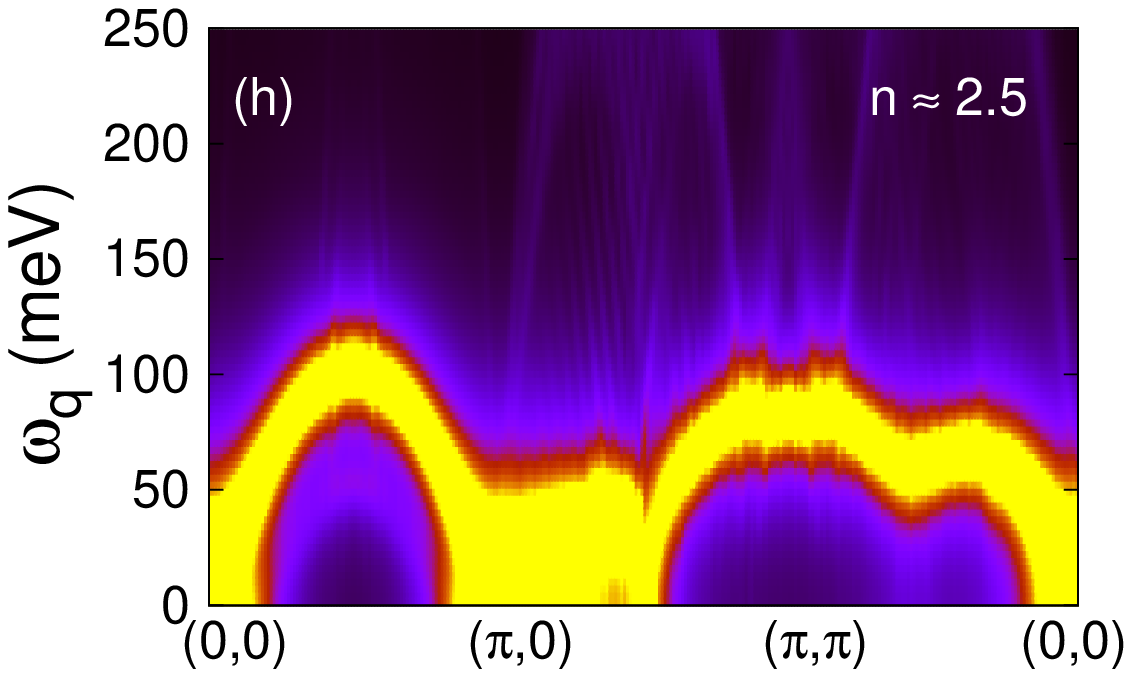}\label{sw4}
\end{minipage}

\vspace{-0.5cm}

\parbox{3cm}{
\hspace{-1.5cm}
\includegraphics[width=0.92\linewidth]{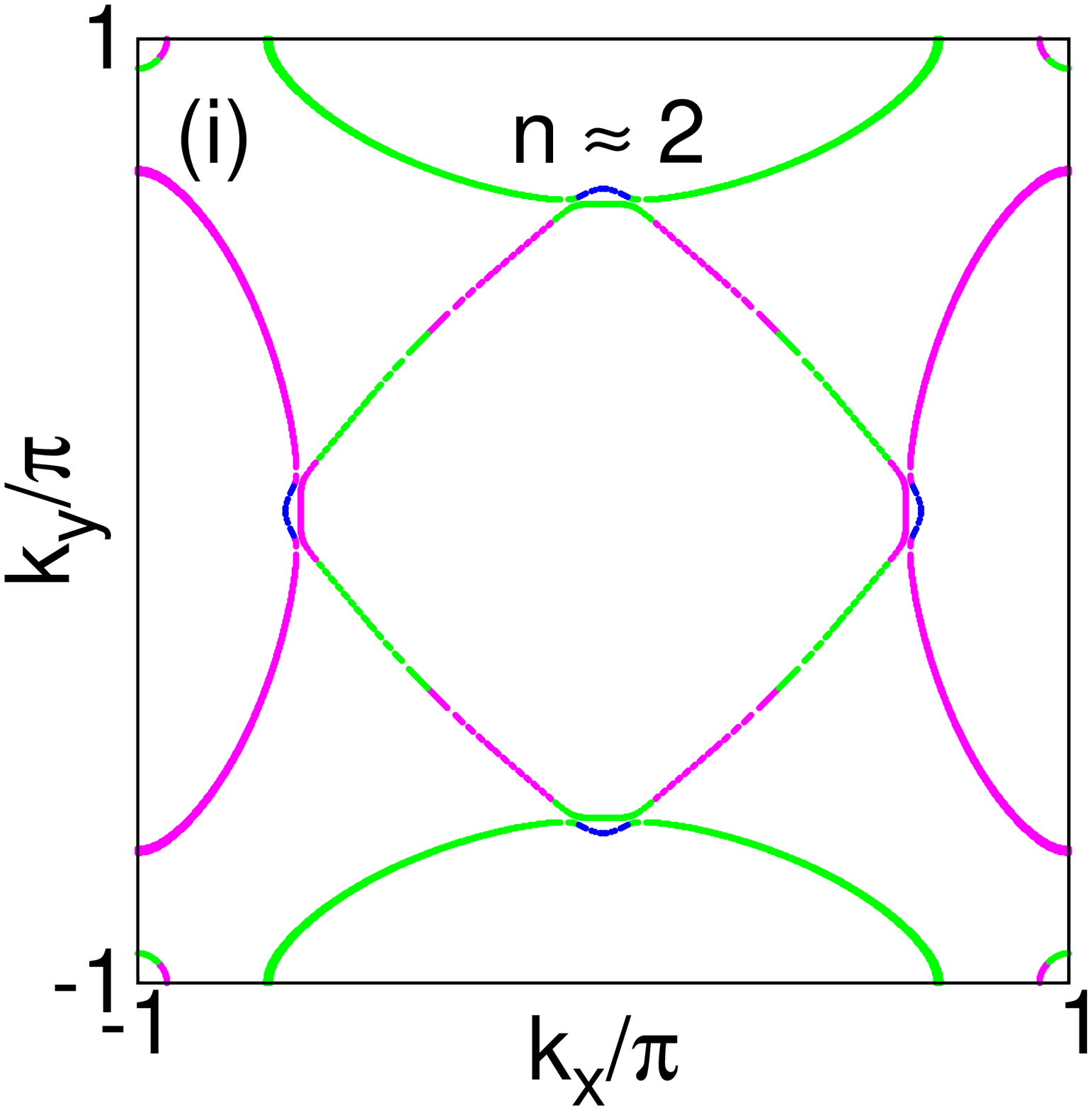}\label{fs5}
\label{fig:2figsF}}
\hspace{-1.0cm}
\begin{minipage}{3cm}
\vspace{-0.1cm}
\includegraphics[width=1.8\linewidth]{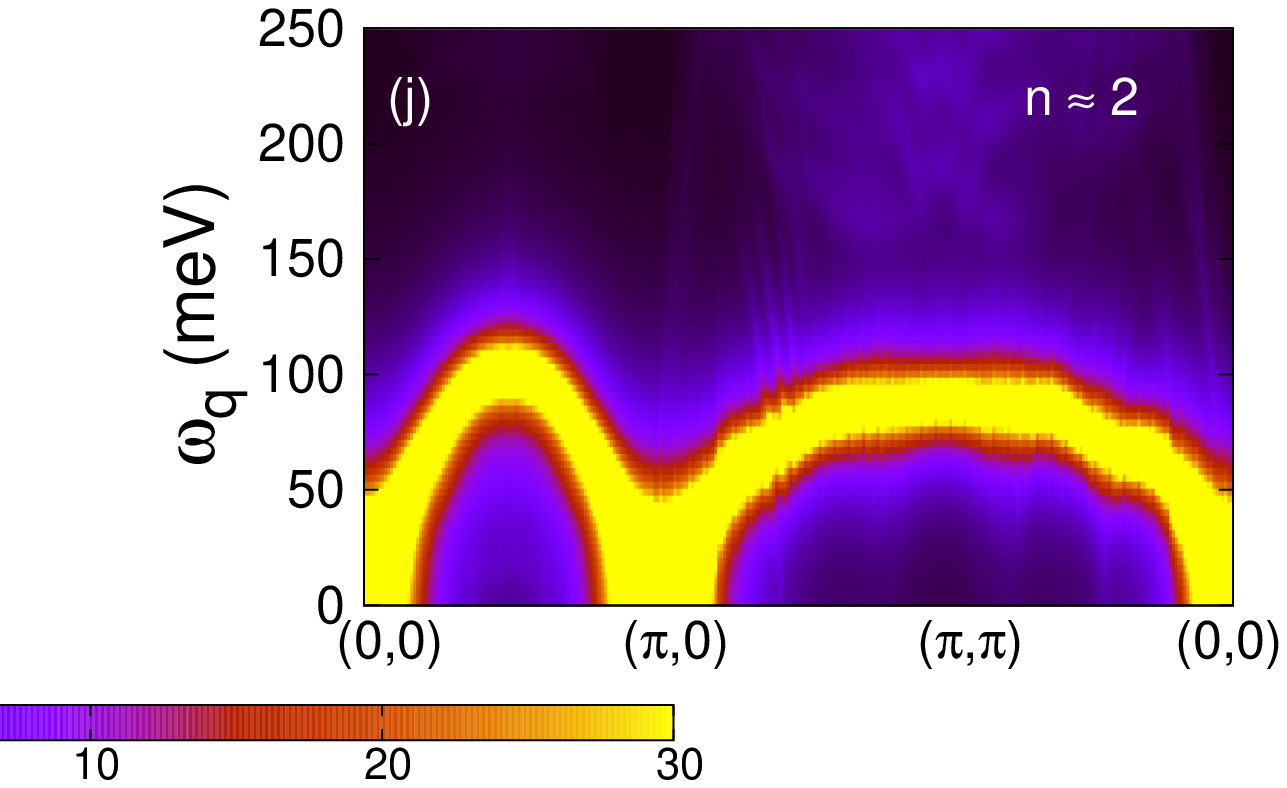}\label{sw5}
\end{minipage}

     \caption{Fermi surface in the PM state (a), (c), (e), (g), (i) and 
             the corresponding spin waves in the $(\pi,0)$ SDW state (b), (d), (f), (h), (j), 
             of the three band model of Ref.~\cite{maria_2010} at various fillings. 
             Contribution of $xz$ (magenta), $yz$ (green) and $xy$ (blue) orbitals in the Fermi surface is indicated. }
    \label{fs_sw_1}%
\end{figure}

In view of this sharply enhanced magnetization for $xy$ orbital upon hole doping, it is of interest to investigate the role of hole doping on spin wave excitations and stability of the SDW state. Fig.~\ref{fs_sw_1} shows the evolution of the spin wave spectral function with increasing hole doping.
Spin wave energy near the AFZB increases rapidly with hole doping and approaches maximum near half-filling (Fig.~\ref{fs_sw_1}f).
The energy at the ferro zone boundary (FZB), on the other hand, decreases on hole doping and vanishes near half-filling, indicating instability of the SDW state. 
Positive spin wave energy at the FZB reappears on further decrease of filling (Fig.~\ref{fs_sw_1}h).
Positive energy modes over whole of the Brillouin zone (BZ) are observed near one-third filling (Fig.~\ref{fs_sw_1}j).
However, the spin wave energy obtained along the F direction is lower than that along the AF direction in contrast to the INS results \cite {zhao_2009}.
The intensity of the spin wave spectral function also increases progressively as the filling decreases from two-third.
The filling of two-third, which is more favourable in the nesting picture, fails to produce a stable $(\pi, 0)$ SDW state. 
Hence the nesting picture alone is not sufficient to account for magnetic excitations.

Although this model does not properly account for the various experimental features of iron pnictides, in order to gain insight into a possible correlation between spin wave stability and the PM state Fermi surface, especially the electron pocket part, the corresponding changes in the latter with hole doping are also shown alongside the respective spin wave spectral functions in Fig.~\ref{fs_sw_1}. The size of the electron pocket reduces with hole doping and becomes diminishingly small near half-filling (Fig.~\ref{fs_sw_1}e). Reappearence of the electron pocket, rotated by $\pi/2$ as compared to the previous cases and significantly larger in size, occurs near one-third filling (Fig.~\ref{fs_sw_1}i). The simulataneous disappearence (near half-filling) and emergence (near one-third filling) of the electron pocket and spin excitation energy highlights the connection between the electron pocket and F spin couplings.

\begin{figure}[H]
\centering
  \includegraphics[width = 3.3in]{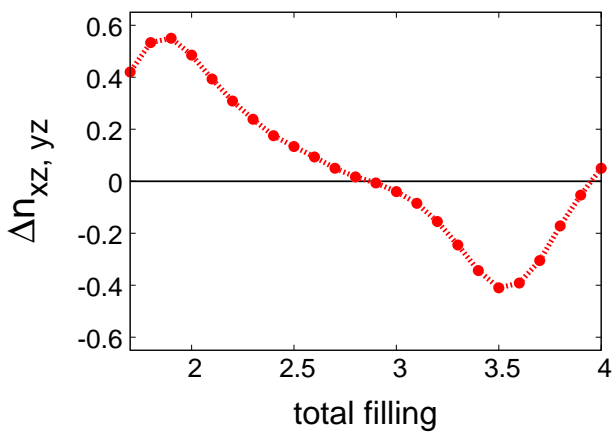}
   \caption{Variation of the orbital order with filling showing sign reversal near half-filling for the three band model of Ref.~\cite{maria_2010}.}
  \label{moo2}
\end{figure}

The orbital order, $\Delta n_{xz,yz}=n_{xz}-n_{yz}$, between $xz$ and $yz$ orbitals is negligibly small near the suggested filling of two-third for this model in the $(\pi, 0)$ state (Fig.~\ref{moo2}). The orbital order starts to develop as the filling decreases from two-third and reaches maxima near $n=3.5$.
However, the calculated orbital order shows an opposite sign as compared to the experimental observations \cite{yi_2011,fuglsang_2011,kim_2013}.
With further decrease in filling, the orbital order changes sign near half-filling and peaks around one-third filling.

Near half-filling, we find that all the densities are close to 1 $(n_{xz}=1.02, n_{yz}=1.04, n_{xy}=0.94)$, hence there is no orbital order. Also, near half-filling, F spin couplings are negligible as they require partially filled bands to be generated (due to exchange of the particle-hole propagator), 
which accounts for the absence of FZB spin wave energy (as shown in Fig.~\ref{fs_sw_1}f). At one-third filling, the $xz$ orbital  density remains close to 1 $(n_{xz}=0.95)$ while the $yz$ and $xy$ bands become partially filled $(n_{yz}, n_{xy}\approx0.5)$.
This accounts for the generation of robust F and AF spin couplings at one-third filling (Fig.~\ref{fs_sw_1}j). Electron/hole doping away from half-filling results in both the orbital order and FZB spin wave energies, thus accounting for the correlation between the two.

\section{Three band model of Ref.~\cite{ghosh_2014_a}}

The hopping parameters for this model \cite{ghosh_2014_a} are:

\begin{table}[h]
\label {tab}
\centering
\begin{tabular}{|c|c|c|c|c|c|c|c|c|}
\hline
$t_1$&$t_2$&$t_3$&$t_4$&$t_5$&$t_6$&$t_7$&$t_8$&$\varepsilon_{xy}$\\
\hline
+0.1&+0.32&-0.29&-0.06&-0.3&-0.16&-0.15&-0.02&0.32\\
\hline
\end{tabular}
\end{table}

\begin{figure}[H]
\centering
\parbox{2.5cm}{
\hspace{-0.7cm}
\includegraphics[width=1.0\linewidth]{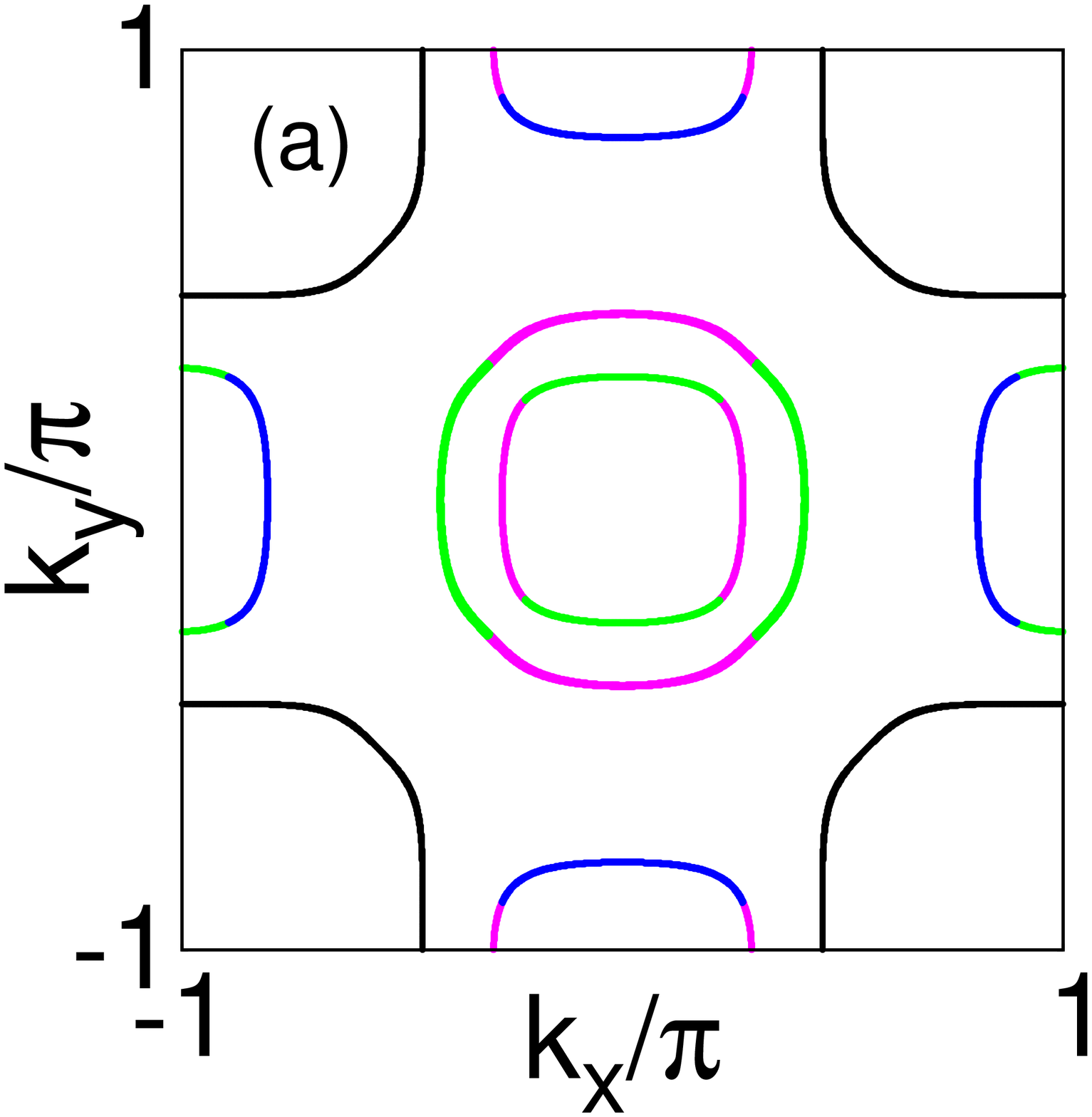}\label{sg_fs}
\label{fig:2figsG}}
\hspace{-0cm}
\begin{minipage}{2.5cm}
\vspace{+0.2cm}
\includegraphics[width=1.6\linewidth]{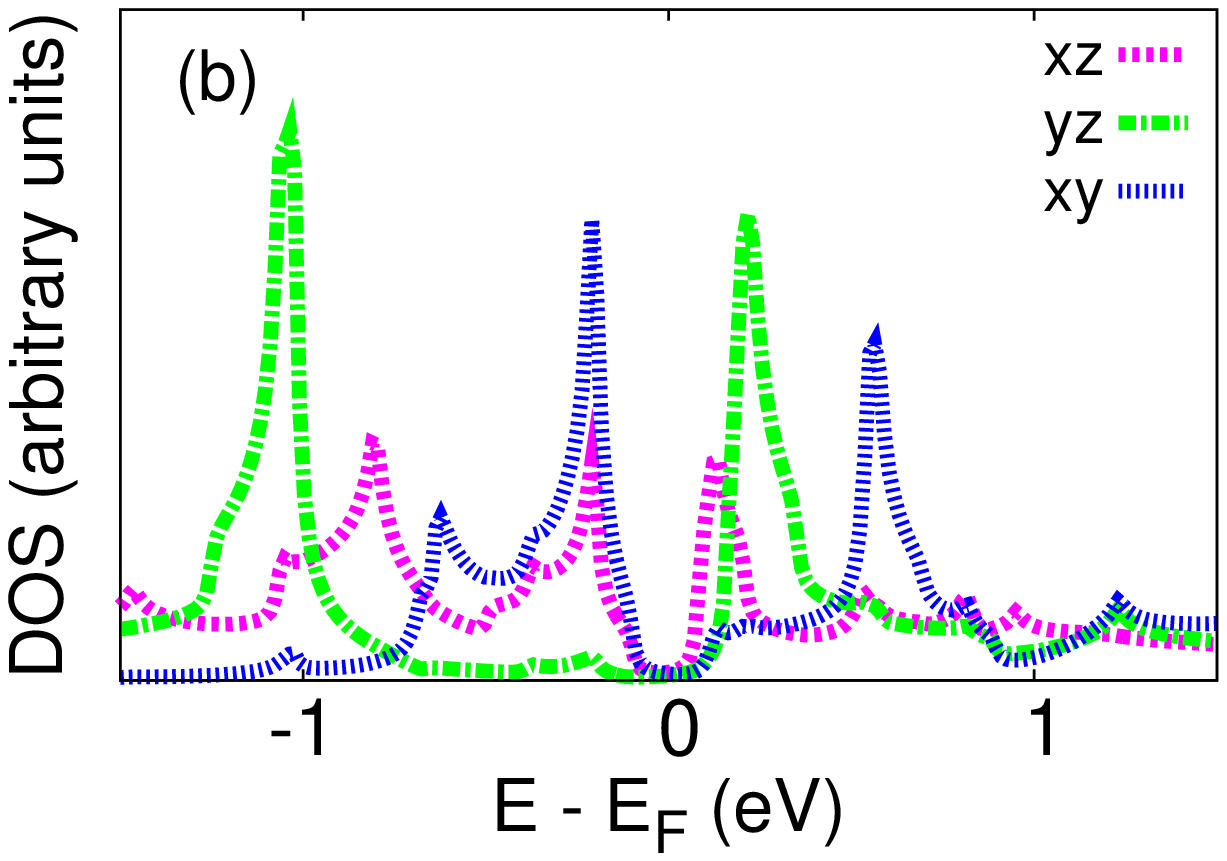}\label{sg_doso}
\end{minipage}
\begin{minipage}{3.5cm}
\includegraphics[width=2.0in]{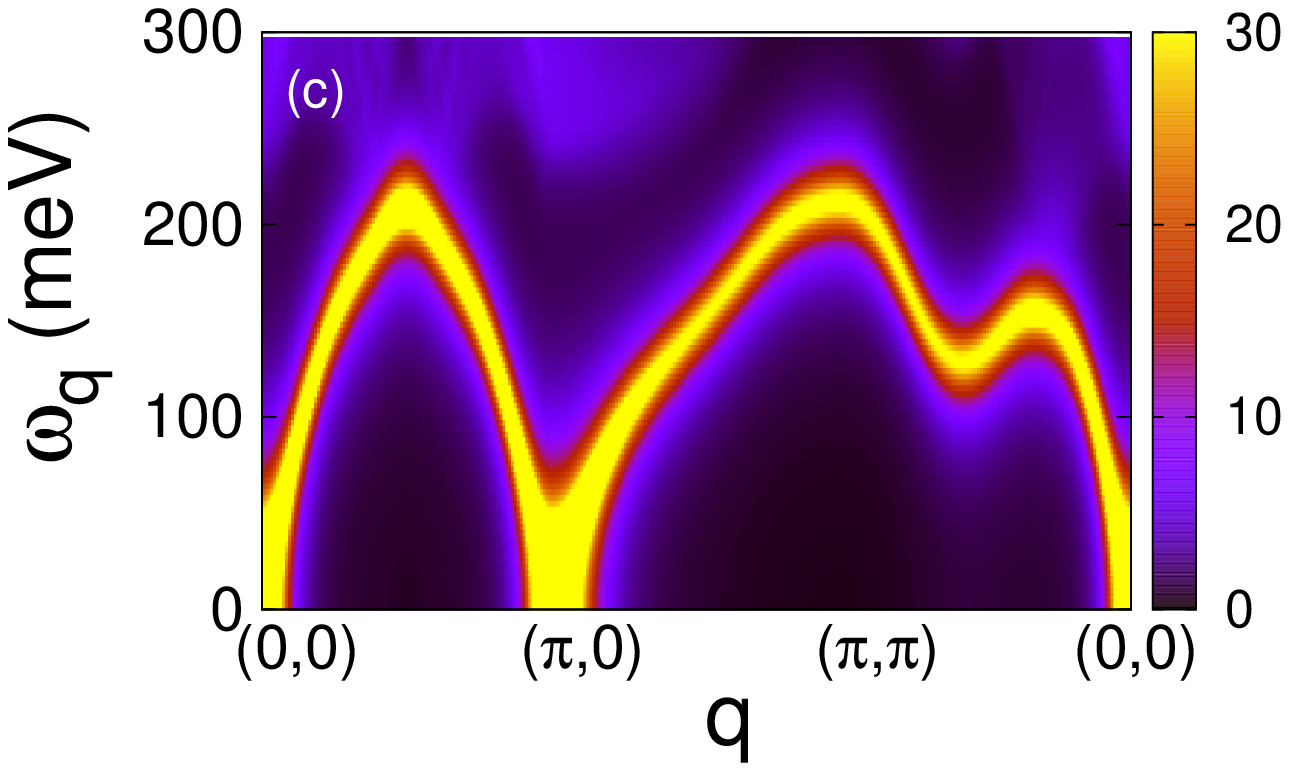}
\end{minipage}

    \caption{(a) Fermi surface in the PM state, 
             (b) Orbitally resolved DOS in the $(\pi,0)$ SDW state and
             (c) Spin wave spectral function showing robust $(\pi, 0)$ SDW state 
                 at half-filling for the three band model of Ref.~\cite{ghosh_2014_a}.} 
    \label{sg_fs_doso_sw}%
\end{figure}

The Fermi surface for this model at half filling has two nearly circular hole pockets of $xz/yz$ character around $\Gamma$ $(0, 0)$ and electron pockets around X $(\pm \pi, 0)$ and Y $(0, \pm \pi)$ points composed of $xy/yz$ and $xy/xz$ orbitals respectively (Fig.~\ref{sg_fs_doso_sw}a). The electron pocket in this model has greater $xy$ share as compared to that in Ref.~\cite {maria_2010}. 
The additional feature near M $(\pi,\pi)$ can be eliminated by including a small third-neighbour term as in the five band model \cite{graser_2010}. The electron/hole pockets and their orbital composition obtained for this model are in good agreement with the ARPES and five band results 
\cite{liu_2009,lu_2009,zhang_2011,kuroki_2008,graser_2009,graser_2010}.

At half-filling, the orbital resolved DOS in the $(\pi,0)$ SDW state (Fig.~\ref{sg_fs_doso_sw}b) shows a gap at the Fermi energy.
This feature is in agreement with DMFT and STM results \cite{yin_2011,chuang_2010}.
Intra-orbital Coulomb interaction is fixed at $U=1.2$ eV for the SDW state analysis of this model.

Stable $(\pi, 0)$ SDW state is observed for this model at half-filling 
(with $n_{xz}=1.2, n_{yz}=1.0, n_{xy}=0.8$), as shown in Fig.~\ref{sg_fs_doso_sw}c.
Strong ferro spin couplings generated for the gapped SDW state are reflected in the high energy spin wave spectrum at the FZB. 
The spin wave spectral function displays closed spin wave structure throughout the BZ, and the energy scale of $200$ meV matches with the INS results 
\cite{zhao_2009,diallo_2009,ewings_2011,harriger_2011}.

\begin{figure}[H]
\centering
  \includegraphics[width = 2.5in]{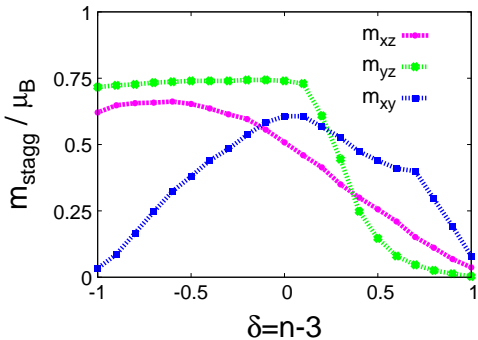}
   \caption{Orbital resolved staggered magnetization as a function of doping with respect to half-filling for the three band model of Ref.~\cite{ghosh_2014_a}, 
             showing well developed moments near half-filling.}
  \label{sg_mvn}
\end{figure}

In order to compare with the model of Ref.~\cite{maria_2010}, we have studied hole/electron doping dependence for this model too.
Staggered magnetization in the $(\pi, 0)$ SDW state drops rapidly on doping away from half-filling, as shown in Fig.~\ref{sg_mvn}.
This behaviour is in accordance with the experimentally observed rapid suppression of Fe moment on doping for iron pnictides \cite{zhao_2008_a}. Orbital resolved staggered magnetization for $yz$ and $xy$ orbitals are negligibly small for the two-third and one-third fillings respectively. Hence, the half-filled SDW state is more robust within this three band model.

\begin{figure}[H]
\centering
  \includegraphics[width = 3.2in]{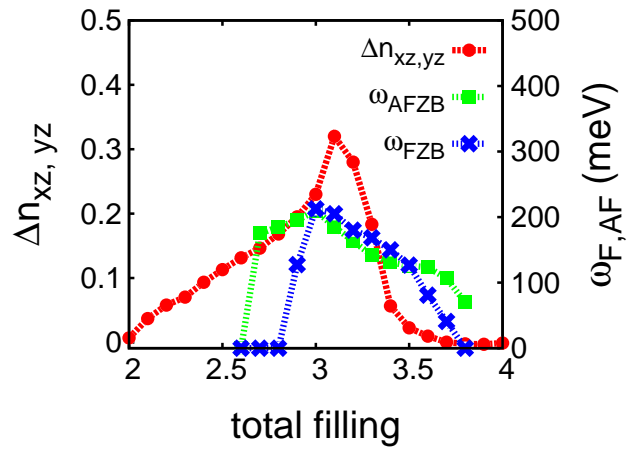}
   \caption{Variation of the orbital order and zone boundary spin wave energies with filling both of which show a maxima near half-filling for the three band model of Ref.~\cite{ghosh_2014_a}.}
  \label{sg_oo}
\end{figure}

The orbital order $\Delta n_{xz,yz}=n_{xz}-n_{yz}$ has a positive sign in the $(\pi, 0)$ state near half-filling, as shown in Fig.~\ref{sg_oo}, which is in agreement with experimental observations \cite{yi_2011,fuglsang_2011,kim_2013}. The orbital order drops on doping with respect to half-filling, which is in agreement with the observation of decrease in splitting between the $xz$ and $yz$ bands on doping in ARPES experiments \cite{yi_2011}. 
The asymmetric behaviour of the calculated orbital order on doping also matches with the experimental observation of rapid drop in the orbital splitting on electron doping as compared to hole doping \cite{yi_2014}. Spin wave energies at the magnetic zone boundaries for various fillings are also shown in Fig.~\ref{sg_oo}. The orbital order as well as the zone-boundary spin wave energies are seen to be maximum near half-filling.

In the $(\pi, 0)$ state, the orbitals $xz$ or $yz$ become half filled depending on which of the two hopping parameters $t_1$ and $t_2$ is dominant. The three band model of Ref.~\cite{maria_2010} has $t_1$ larger than $t_2$, inducing the $xz$ orbital to become half filled ($n_{xz}\approx 1$) with a saturated staggered magnetization over a broad range of fillings (Fig.~\ref{m_mvn}). As filling increases beyond half filling, occupation in the $yz$ orbital exceeds that in $xz$ (Fig. 5), reversing the sign of the orbital order. In contrast, the hopping parameter $t_2$ is dominant in the three band model of Ref.~\cite{ghosh_2014_a}, resulting in half filling and saturated staggered magnetization for the $yz$ orbital ($n_{yz}\approx 1$) near and below half filling (Fig.~\ref{sg_mvn}). As seen in Fig.~\ref{sg_oo}, the $xz$ orbital density grows progressively with filling ($n_{xz} > 1$) and sharply peaks when total filling is near 3, thereby giving the correct sign of the orbital order at half filling.

The orbital matrix components of the transverse spin fluctuation spectral weight in the SDW state 
have been investigated recently in a five-band model for iron pnictides \cite{kovacic_2015}. 
The two orbitals $(xz,x^2-y^2)$ which yield significantly lower spectral weight compared to the other three orbitals $(yz,xy,3z^2-r^2)$, 
are also overfilled with more than average filling of 1.2 per orbital and have minimum staggered magnetization.
On the other hand, the orbitals $(yz,xy)$ which are half filled yield the maximum staggered magnetization.
Interestingly, a single set of orbital correspondence: 
$xz \Leftrightarrow (xz,x^{2}-y^{2})$, $yz \Leftrightarrow (yz,xy)$ and $xy \Leftrightarrow 3z^{2}-r^{2}$ 
between the three band and five band models is consistent for all the three orbital resolved magnetic properties of the SDW state mentioned above, 
each of which involve band summations and not just the states at the Fermi surface.
In the three band model, the staggered magnetization near half filling (Fig.~\ref{sg_mvn}) is highest for $yz$ (which is also half filled), lowest for $xz$ (which is also overfilled), and the $(xy,yz)$ orbitals exhibit maximum spin wave spectral weight. 
The strong hybridization of the $3z^2-r^2$ orbital with $x^{2}-y^{2}$ orbital \cite{kovacic_2015} is also
consistent with the above correspondence.
  
\section{Conclusion}

A comparative study has been carried out in this paper focussing on orbital resolved magnetizations, spin wave excitations, orbital order, and their doping dependence in the $(\pi,0)$ SDW state for two different three band models. The three band model at two-third filling \cite{maria_2010} yields qualitatively correct Fermi surface structure with two circular hole pockets around $(0,0)$ and elliptical electron pockets near $(\pi,0)$ and $(0,\pi)$.
However, the calculated spin wave spectral function in the $(\pi, 0)$ SDW state shows negative spin wave energies near the AF zone boundary, indicating that the magnetic state is unstable.
The main conclusion of this paper, as inferred from the sharp enhancement of magnetic moments, spectral intensity, and spin wave energies upon hole doping, is that the magnetic state instability at $n=4$ is associated with weakly developed magnetic moments due to overfilling resulting in weak magnetic couplings and low excitation energies.

Furthermore, in the $(\pi, 0)$ SDW state for this model \cite{maria_2010}, the negative sign of the orbital order $n_{xz} - n_{yz}$ near two-third filling is not in agreement with experiments. Orbital order is a composite effect of the anisotropic magnetic ordering and anisotropic hopping.
The negative sign of the orbital order follows directly from the choice of the first two hopping parameters $t_1 > t_2$ in this model, which differs from that ($t_2 > t_1$) obtained in the tight-binding fit of DFT calculations \cite{lee_2009,graser_2010}.

In contrast, the three band model at half filling \cite{ghosh_2014_a}, with two hole pockets around $\Gamma$ and electron pockets near X and Y points in agreement with DFT results, yields well developed magnetic moments in the $(\pi, 0)$ SDW state. The gapped SDW state yields magnetic excitations in excellent agreement with the INS experiments. The anisotropic suppression of the calculated orbital order with hole and electron doping is also in agreement with ARPES experiments. Reduction of magnetic excitation energy scales with both hole and electron doping also indicates the half-filled SDW state to be more appropriate within the three band sector. The sign of the orbital order in this model \cite{ghosh_2014_a} near half filling ($n_{xz} - n_{yz} \approx +0.2$) is also in agreement with experiments.

Both the zone boundary spin wave energies in the F and AF directions as well as the orbital order simultaneously peak near half filling, highlighting the correlation between orbital order and SDW state stabilization
\cite{yi_2009,lee_2009,lv_2009,kruger_2009,chen_2010,lv_2010,yanagi_2010,lv_2011,applegate_2012,luo_2013,fernandes_2014}. The strong enhancement of $\omega_{\rm FZB}$ due to partially filled $xy$ band strongly supports the origin of F spin couplings due to exchange of particle-hole propagator as in metallic ferromagnets.

Finally, we summarize few broad conditions on the microscopic Hamiltonian parameters which favor stabilization of the $(\pi,0)$ SDW state within the constraints on the Fermi surface structure and the minimal three band model involving only $xz$, $yz$ and $xy$ orbitals. i) The sum $t_1 + t_2$ should be positive to avoid overfilling by pushing up the $(\pi,\pi)$ state band energy in relation to the $(0,0)$ state. ii) The condition $t_2 > t_1$ ensures correct sign of orbital order when the $xz-yz$ sector is more than half-filled. iii) For strong ferro spin couplings to be generated, the $xy$ orbital should be less than half-filled. iv) Disparately large band width should be avoided as it results in weak local moments in the SDW state, as for the $xy$ orbital in Ref.~\cite{maria_2010} where the hopping terms $t_5$ and $t_6$ are an order of magnitude larger than other hopping terms.

\end{document}